\documentclass[12pt, a4paper, titlepage]{article}
\usepackage{graphicx}
\title{Absolute calibration of an EMCCD camera by quantum correlation linking photon counting to analog regime}
\author{A. Avella, I. Ruo Berchera, I. P. Degiovanni, G. Brida and	M. Genovese\\
\\
	INRIM, Strada delle Cacce 91, I-10135 Torino, Italy  \\
}

\date{\today} 
\begin{document}
\maketitle

\begin{abstract}
We show how the same set-up and procedure, exploiting spatially multi-mode quantum correlations, allows the absolute calibration of a EMCCD camera from the analog regime down to the single photon counting level, just by adjusting the brightness of the quantum source. At single photon level EMCCD can be operated as an on-off detector, where quantum efficiency depends on the discriminating threshold. We develop a simple model to explain the connection of the two different regime demonstrating that the efficiency estimated in the analog (bright) regime allows to accurately predict the detector behaviour in the photo-counting regime and vice-versa. This work establishes a bridge between two regions of the optical measurements that up to now have been based on completely different standards, detectors and measurement techniques. 
\end{abstract}


One of the main direction in modern optics is the development of quantum technologies, such as quantum communication, imaging and sensing, where single photon \cite{pin1,in3,pin2,pin3} or few photon states of light are generated manipulated and detected \cite{in1,in2,in4,geno}. Development of dedicated methods for characterizing detectors in this context is necessary, as widely recognised inside the radiometric community \cite{qcan1}.
Single photon regime is far from the one where traditional radiometry operates and where the best accuracy is available \cite{fox}. Some specific activities in this context are already on-going\cite{qcan2}, in particular related to the calibration of single-photon detectors exploiting the Klyshko's twin-photon coincidence technique \cite{bp1,b,a,b1,Brida2000a, PolMig2007, Cheung2011, Polyakov2009} and its developments \cite{haderka,n, pe,aga,s,v, Worsley2009, Schmunk2011}.

Beside the commonly used on-off and single-photon avalanche photodiodes without spatial resolution new types of detectors are considered for specific application to overcome their limitation, namely photon number resolving and spatially resolving detectors. 
Among them Electron-multiplication charge-coupled-devices (EMCCDs) cameras
 represent a commercial and diffuse approach for single- and few-photon imaging when high spatial resolution is required. EMCCDs and intensified CCD have been used also for top level experiments in quantum optics and quantum information technology when many spatial modes have to be detected, from sub-shot noise
imaging \cite{BridaNP2010}, quantum illumination \cite{qm2} and ghost imaging \cite{Meda2015} to detection of EPR state and entanglement of orbital angular momentum \cite{BlanchetPRL2008,EdgarNC2012,FicklerSR2013,AspdenNJP2013,JachuraOL2015}.

In this paper we present the absolute calibration of a EMCCD camera, operated in photon counting regime as a threshold detector ("on-off") pixel by pixel, by the measurement of spatially multi-mode quantum correlation in squeezed vacuum. The "on-off" behaviour is achieved by applying a discriminating threshold $T$ on the electron counts $n_{e}$ at each pixel: a photon is detected if $n_{e} > T$. In this regime, the camera works as a non-linear photon number resolving detector, counting the number of incident photons in a region of many pixel (spatial multiplexing) and acquiring many frames (time multiplexing). Hereinafter, we assume to work in condition of low illumination (negligible probability to have more than one photon per pixel per frame); in such condition the device can be approximated as a linear photon counting detector. 

The main difference between the Klyshko's twin photon coincidence technique and the method presented here  stems from the fact that we compared the number of detected photons in correlated areas in a large integration time, exploiting a technique that was developed for CCD in analog regime \cite{BridaJOSAB2006, BridaOE2008, BridaJMO2009, RuoASL2009, Brida2010, Meda2014}. 
In particular, we measure the noise reduction factor $\zeta = \langle[\delta(N_1-\alpha N_2)]^2\rangle/ \langle  N_1- \alpha N_2)\rangle$  and the correlation $C = \langle N_1 N_2 \rangle - \langle N_1\rangle \langle N_2\rangle $; where $\alpha=\langle N_1\rangle /\langle N_2\rangle$, $N_1$ and $N_2$ being the number of photons in the two correlated areas. These quantities are related to the mean detection efficiency $\eta$ by: $\zeta  \simeq \frac{1+\alpha}{2}- \eta A$ and $C \simeq \eta A \langle N_1 \rangle$ both in analog and photon counting regime for low photon number per mode; where $A$ is a geometrical parameter \cite{Meda2014}. 

Even if, in principle, also in the case of an EMCCD operated in photon counting it would be possible to exploit the Klyshko's twin photon technique, there are a certain number of practical reasons preventing its use: Klyshko's technique needs an illumination regime providing not more than one coincidence per frame, but in this illumination range the noise becomes dominant. Moreover, the read time of an EMCCD is higher respect to single photon detectors, yielding Klysko's technique very slow. 

It is important to note that the detection efficiency includes both the transmission efficiency of the optical detection system and the CCD camera quantum efficiency. The efficiency of the CCD could be measured by performing a conventional calibration of the losses in the optical channel through a trasmissivity measurement (as in the Klyshko's twin-photon technique). 

In this work, we exploit the same procedure for the absolute calibration of the detection efficiency both for the EMCCD operating in proportional (analog) regime (i.e. without electron-multiplication) and when it is operating in photon counting regime under the assumption of low illumination. Then, we compare the results obtained in the two regimes for the same device and the same experimental set-up (only the power of the source of twin-photon has to be tuned, as well as the mode of operating of the camera). We have to consider that in photon counting regime the detection efficiency is a function of the threshold. Therefore, it is not possible performing a direct comparison between the efficiencies for the two regimes. However, it is possible calculating the dependence $\eta(T)$ by the measured value of the analog quantum efficiency $\eta_0$ and to compare this with the measured efficiency in photon counting. 

This quantum efficiency measurements, obtained for the same detector operated in proportional and photon-counting regimes, is the most significant result of this paper that takes advantage of the unique versatility in terms of regimes of operation of our detector.


In order to obtain the theoretical behaviour for the noise and for the quantum efficiency $\eta(T)$ of our camera, we analyse the typical model of an EMCCD \cite{1,2}, then we estimate the parameters involved in this model by means of a set of measurements that are independent on the absolute calibration.

For $n$ photoelectrons at its input, the multiplication stage of each pixel provides a random number of electron counts $x$ following the distribution \cite{Lantz2008}: 
\begin{eqnarray}
\mathcal{P}(x|n) =\frac{ x^{n-1} exp(-x/g) }{ g^{n}(n-1)!} \;\;\;\;\; for \;\;\; n>0; \label{1}\\
\mathcal{P}(x|n) = \delta(x)  \;\;\;\;\; for \;\;\; n=0; \label{2}
\end{eqnarray}
where g is the mean multiplication gain.
As the total number of electron counts per pixel is due to the contribution of photoelectrons multiplication counts and to the noise counts, the electron counts random distribution at the output is the convolution of $\mathcal{P}(x|n)$ with the noise distribution:  
\begin{equation}
P_{tot}(x|n) = \int^{\infty}_{-\infty} \mathcal{P}(y|n)  P_{noise}(x-y) dy.
\end{equation}

The most relevant noise contributions typically affecting an EMCCD are: read noise, dark current and spurious charges. 

The read noise, generated by the on-chip output amplifier, follows a Gaussian distribution $P_{rn}(x; \mu,\sigma)$ where the mean value $\mu$ is the bias level of the read-out distribution and the standard deviation $\sigma$ characterizes the fluctuation of read noise:
\begin{equation}
P_{rn}(x; \mu,\sigma) = \frac{1}{\sigma \sqrt{2 \pi}} exp \left[ \frac{-(x - \mu)^{2}}  {(2 \sigma^2)} \right]. \label{gauss}
\end{equation}

The dark current is due to thermally generated charges and strongly depends on temperature and acquisition time. Generally, it can be independently measured and subtracted from data. In our case, the camera parameters can be set to have a negligible contribution of dark current.

Spurious charges, also called Clock Induced Charge (CIC), are created during the fast clock variations required for shifting the photoelectrons to the readout register. CIC generate an electron counts distribution  $P_{sc}(x|n)$ that has the same behaviour of equations \ref{1} and \ref{2}, but with gain $g_{sc}$ lower than $g$.

In a reasonable operating configuration, the probability to have more then one spurious charge is negligible. Therefore, the random probability distribution of the electron counts, in the absence of illumination (i.e. electrons generated by photon absorption), is:
\begin{equation}
P_{noise}(x)= (1-p_{sc}) P_{rn}(x) 
+ p_{sc} \int_{-\infty}^{\infty}P_{rn}(y) P_{sc}(x-y|n=1) dy,
\label{nos}
\end{equation}
where 
$p_{sc}$ is the probability to have a spurious event,
and the probability that a pixel clicks due to the noise is: 
\begin{equation}
Noise(T) = \int_T^{\infty} P_{noise}(x) dx.
\label{nos2}
\end{equation}



The definition of detection efficiency, for a camera in single photon counting regime, is: $\eta(T)  =  P_{true}( T)/p_{ph},$
where $P_{true}(T)$ is the probability that a pixel clicks ($x \geq T$) due to an incident photon and $p_{ph}$ is the probability to have an incident photon.
When operating with a sufficient low light level, it is possible to assume that at most one photon is detected per pixel. Under this condition, the probability that a pixel clicks is:
\begin{equation}
P_{click} \simeq \eta_0 p_{ph} P_1(x \geq T) + (1-p_{ph}) P_{noise}(x \geq T)
\end{equation}
where $\eta_0$ is the probability that an incident photon generates a photo-electron and corresponds to the analog detection efficiency. 
If we use a threshold sufficiently high to cut out the main part of the read noise ($T \geq 2 \sigma +\mu$), it is possible to assume that double detection event on the same pixel are negligible (for double detection events we intend all possible combinations: photon-photon, photon-noise, noise-noise). Therefore, we have:
\begin{eqnarray}
P_{click} \simeq \eta_0 p_{ph} P_1(x \geq T) +  P_{noise}(x \geq T) \\
 P_{true}(x \geq T) = \eta_0 p_{ph} P_1(x \geq T) \Rightarrow  \eta(T) = \eta_0  P_1(x \geq T).
 \label{ref}
\end{eqnarray}
hence, the number of click per frame is: $N_{click} = N_{true} + N_{noise}$ and we can measure the number of true counts, used for calculate $\zeta$ and $C$, as $N_{true} = N_{click} - N_{noise}$.

The function $P_1(x)$ represents the electron counts distribution under the assumption that there is at most one photon per pixel, that there are no spurious counts and taking into account the contribution of the read noise:
\begin{equation}
P_1 = \int_{-\infty}^{\infty} \frac{1}{\sigma \sqrt{2 \pi}} e^{\frac{{-(y-\mu)^2}}{2 \sigma^2}} \mathcal{P}_1(x-y) dy,
\label{P1}
\end{equation}
where $ \mathcal{P}_1(x) = \mathcal{P}(x|n=1) = g^{-1} e^{-x/g}$
is the distribution of electron counts generated by one photon.





For what concerns our camera, we have estimated independently all the parameters involved in the models. Therefore, we are able to predict the behaviour of detection efficiency and noise in function of an arbitrary threshold $T$ and in function of the analog detection efficiency $\eta_0$.

\begin{figure}[h]
\centering
\includegraphics[width=8cm]{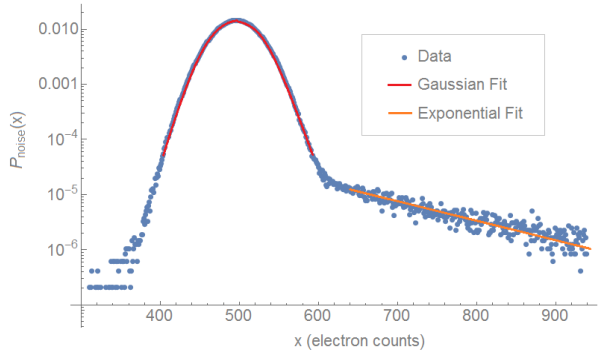}
\caption{ Logarithmic scale histogram of the electron counts distribution for a frame acquired without incident light (blue dots). 
Gaussian (red line) and  
exponential (orange line).}
\label{fig:histo}
\end{figure}

Figure \ref{fig:histo} presents the histogram of the electron counts distribution for a frame acquired with closed shutter (no input light): showing an excellent agreement between the experimental data and the theoretical prediction in equation \ref{nos}. Two different behaviours are clearly visible: at low counts level the Gaussian contribution of read noise is dominant; instead, at high counts level only the exponential contribution of spurious charges is observable. 
By fitting the theoretical distributions of read noise, on the first part of the histogram we estimate  $\mu=507.9(0.3) counts/pixel$, $\sigma= 24.88(0.03) counts/pixel$ and by fitting the distribution of spurious charge on the second part of the histogram we estimate the parameters $p_{sc}=0.0044(0.0006)$ and $g_{sc} = 141(2) counts/pixel$.

\begin{figure}[h]
\centering
\includegraphics[width=8cm]{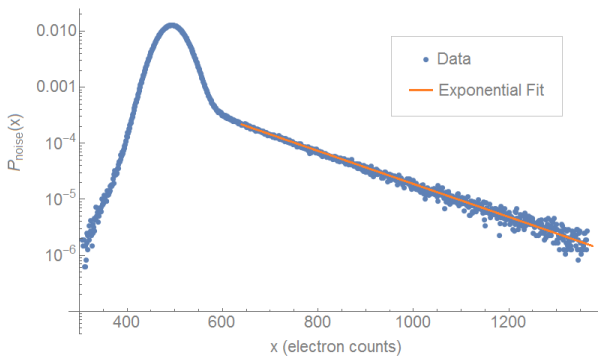}
\caption{Logarithmic scale histogram of the electron counts distribution for a frame acquired with incident light (blue dots). 
Exponential curve fit (orange line).} 
\label{fig:histo2}
\end{figure}

Figure \ref{fig:histo2} shows the histogram of the electron counts distribution for a frame acquired  with input light.
The light intensity was selected to avoid double events, but, at the same time, to guaranty that spurious charges are negligible respect to photoelectrons. 
By fitting the theoretical distributions $\mathcal{P}_1(x)$ on the part of the histogram in which the read noise is negligible we estimate the mean multiplication gain of the camera $g = 147(2)counts/pixel$.

 All the uncertainties are evaluated on a set of 35 frames acquired under the same illumination conditions. For each frame we perform a fit. Figures \ref{fig:histo} and \ref{fig:histo2} are examples of two particular frames histograms, in presence and in absence of illumination respectively.

\begin{figure}[htbp]
\centering
\includegraphics[width=7cm]{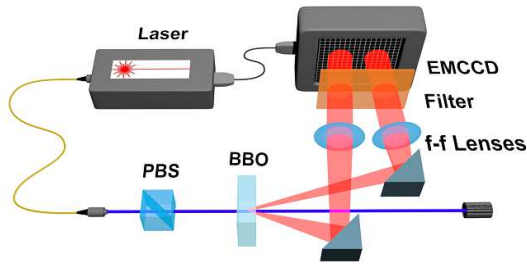}
\caption{Schematic representation of the experimental apparatus.}
\label{fig:setup}
\end{figure}

Our set-up (Figure \ref{fig:setup}) is composed by a diode laser, operating at $\lambda = 406 nm$ in pulsed mode, synchronized with the exposure time of a electro-multiplied CCD camera (EMCCD). The laser is coupled in a single mode fibre providing a well Gaussian shaped beam, the beam is then collimated, and a polarising beam splitter (PBS) selects the vertical component of the polarization. The laser beam pumps a $5 \times 5 \times 15$ $mm^3$ BBO non-linear crystal of Type II, where two correlated beams are produced through spontaneous parametric down conversion. The generated twin beams are sent to the EMCCD by two plane mirrors. A far field lens with a focal length of $f = 10$ $cm$ is located in a $f-f$ configuration, between the output surface of the crystal and the detection plane.
An interference filter at 800 nm with 40 nm bandwidth and a central transmittivity of $99\%$ is put in front of the camera. The phase matching is set to maximizing the emission around the degenerate wavelength ($\lambda = 812 nm$).

\begin{figure}[h]
\centering
\includegraphics[width=8cm]{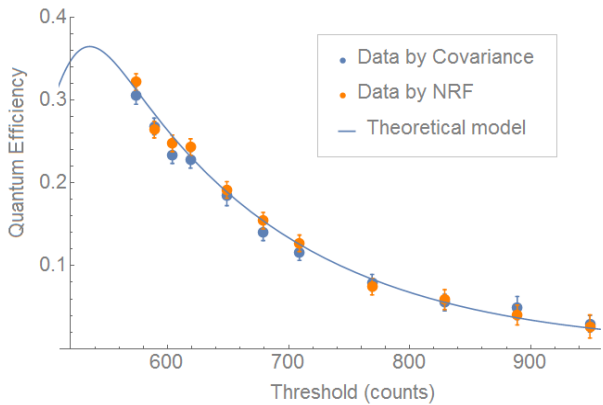}
\caption{Quantum efficiency, in function of the threshold $T$, estimated exploiting spatial correlations of quantum twin beams. The continuous line shows the theoretical prediction calculated by equations \ref{ref} and \ref{P1}, and by mean of the preliminary estimation of $\eta_0$.}
\label{fig1}
\end{figure}

\begin{figure}[h]
\centering
\includegraphics[width=8cm]{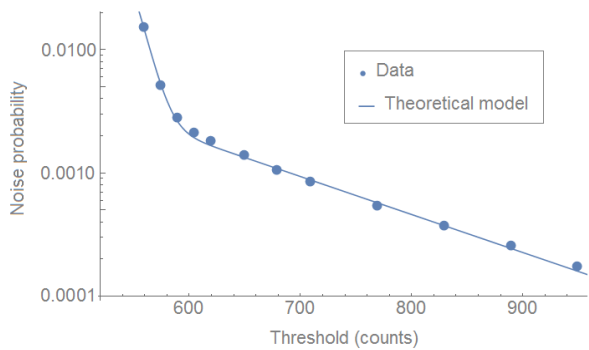}
\caption{Measured probability to have a click per pixel per frame due to a noise event in function of the threshold (uncertainty bars are smaller than dots). Continuous line represents the prediction obtained by our model: equations \ref{nos} and \ref{nos2}.}
\label{fig2}
\end{figure}

Figure \ref{fig1} shows the behaviour of $\eta(T)$ derived by the model of EMCCD (Equation \ref{ref}) and compares it with the experimental results showing a perfect agreement with the theoretical prediction. The analog quantum efficiency, derived using the same technique based on twin beam in analog regime\cite{Brida2010, Meda2014}, is $\eta_0 = 0.54(0.02)$. It is worth recalling that $\eta_0$ and $\eta$ represent the detection efficiency of the beam at $812nm$, including both the transmission efficiency of the optical detection system and the CCD camera quantum efficiency.
It is important to note that the model and the measurement technique, based on twin beams, are not valid for low threshold level when the noise contribution becomes dominant. In our system this happen approximately for $T < 560 \; counts$. 
Figure \ref{fig2} shows the level of noise in photon counting regime and the corresponding theoretical level provided by the Equation \ref{nos}.

The results shown in Figure \ref{fig1} represent the first absolute calibration of the quantum efficiency of an EMCCD operating as a threshold detector. 

The comparison shows that the behaviour of the detection efficiency of the EMCCD, in on-off regime after threshold application, is completely predictable by a simple model that is function of the measured value of the gain of the electro-multiplication register, of the efficiency measured in the analog regime and of the chosen threshold $T$.
Thus, on the one side we achieve a complete characterization and understanding of the behaviour of a device acquiring more and more importance in the field of single photon measurement. On the other side, we establish a bridge between the light intensity level typical of classical radiometric measurements, the analog regime and the quantum radiometry operating at single-photon level \cite{qcan1,qcan2}. 

The possibility to use the same absolute calibration technique, for the same device, both for photon counting regime and for analog regime provides a radiometric link between low illumination regime (few photons) to the mesoscopic and to the classical macroscopic ones. This represents an important step in the development of quantum radiometry providing traceability of measurement at the few photon level, the relevant illumination level for most of the emerging quantum technologies.

\section*{Funding Information}
This work has received funding from the projects EU-FP7 BRISQ2, EMRP (EXL02 - SIQUTE), (14IND05 MIQC2 project), and from FIRB No. D11J11000450001 (MIUR).







\end{document}